\documentclass[iop]{emulateapj}
\usepackage{epsfig,natbib}
\usepackage{graphicx}
\received{2015 June 2} 
\accepted{2015 August 20} 
\slugcomment{Accepted by AJ}

\newcommand{\eke}{\emph{Kepler\ }}
\newcommand{\ek}{\emph{Kepler}}

\shortauthors{Gilliland {\it et~al.\/}}
\shorttitle{Kepler Noise Properties}

\begin{document}
\pagenumbering{arabic}

\title{Kepler Mission Stellar and Instrument Noise Properties Revisited}
\author{Ronald~L.~Gilliland\altaffilmark{1,2}, 
William~J.~Chaplin\altaffilmark{3,4}, 
Jon~M.~Jenkins\altaffilmark{5}, 
Lawrence~W.~Ramsey\altaffilmark{1}, 
Jeffrey~C.~Smith\altaffilmark{5,6} 
}
\altaffiltext{1}{Department of Astronomy and Astrophysics, and Center for Exoplanets and Habitable Worlds,
The Pennsylvania State University, 525 Davey Lab, University Park, PA 16802, USA; gillil@stsci.edu}
\altaffiltext{2}{Space Telescope Science Institute, 3700 San Martin Dr., 
Baltimore, MD 21218, USA}
\altaffiltext{3}{School of Physics and Astronomy, University of Birmingham,
Edgbaston, Birmingham, B15 2TT, UK}
\altaffiltext{4}{Stellar Astrophysics Centre (SAC), Department of Physics
and Astronomy, Aarhus University, Ny Munkegade 120, DK-8000 Aarhus C, Denmark}
\altaffiltext{5}{NASA Ames Research Center, Moffett Field, CA 95035, USA}
\altaffiltext{6}{SETI Institute, 189 Bernado Ave., Suite 100, 
Mountain View, CA 94043}

\begin{abstract}
An earlier study of the {\em Kepler Mission} noise properties on time
scales of primary relevance to detection of exoplanet transits found
that higher than expected noise followed to a large extent from the
stars, rather than instrument or data analysis performance.
The earlier study over the first six quarters of {\em Kepler} data is
extended to the full four years ultimately comprising the mission.
Efforts to improve the pipeline data analysis have been successful in
reducing noise levels modestly as evidenced by smaller values derived from the 
current data products.  The new analyses of noise properties on transit
time scales show significant changes in the component attributed to 
instrument and data analysis, with essentially no change in the inferred
stellar noise.  We also extend the analyses to time scales of several days,
instead of several hours to better sample stellar noise that follows from
magnetic activity.  On the longer time scale there is a shift in stellar
noise for solar-type stars to smaller values in comparison to solar values.
\end{abstract}



\section{Introduction}

The NASA {\em Kepler Mission} has left an indelible imprint on 
exoplanet and stellar properties research through its unmatched 
combination of photometric precision for a large number of stars
($\sim$150,000), over a long period of time (4 years) with a 
standard observing cadence of 30 minutes \citep{koch10}.
The exquisite time series returned from \eke have provided the
first results for Earth-sized planets potentially in or near
the habitable zones of their host stars \citep[e.g.,][]{boru13,torr15}.
The standard cadence data have revolutionized
our ability to probe the properties of red giants with
asteroseismology \citep{bedd11}, while the limited short-cadence,
1 minute observations have similarly revolutionized asteroseismology
of dwarf stars \citep{chap11}.

While \eke photometric time series are excellent compared to anything
previously available, they are not perfect and one of the early 
surprises in the {\em Kepler Mission} was a higher than expected
noise level, CDPP -- Combined Differential Photometric Precision \citep{chri12a},
a roll-up of all factors of relevance for detection of exoplanet
transits with widths of 3 -- 12 hours.  The {\em Kepler Mission} had
been designed \citep{koch10} to have roughly comparable noise levels
for fiducial 12th magnitude solar-type stars arising from irreducible
Poisson fluctuations, and intrinsic noise from the stars, with smaller
contributions expected from imperfections in the instrument, and the
software used to provide extracted and calibrated time series.
\eke provided the first opportunity to observe stars other than the 
Sun at precision levels allowing well informed inferences about the
intrinsic variations of solar-type stars.  The total noise (CDPP
at the nominal 6.5 hours) was found to most commonly be 30 parts per
million (ppm) for 12th magnitude solar-type stars, compared to an
expected 20 ppm \citep{jenk02}.  This higher than expected noise level resulted in
the need for twice the data extent to reach the original mission goals,
and was a prime motivation in seeking to extend the original 3.5 year
mission.  An extended mission was approved to double the original
extent, however the loss of two (of four) reaction wheels brought
the prime mission to an end after rather precisely 4 years of 
observing.  Analyses by \citet{gill11} showed that the primary factor
in increased CDPP was the contribution from stars, with 
a smaller addition from imperfections of instrument and software.

Several studies have addressed general stellar variability with 
\eke data.  \citet{ciar11} presented an overview of variability from
the first month of data over most stellar types.  \citet{mcqu12}
and \citet{robe13}
also analyzed the first month.  \citet{basr13} and \citet{walk13} 
used one quarter of data to focus on
a multi-time scale consideration of solar-type stellar variability 
concluding that the \eke stellar sample tended to be quieter than
the average Sun, a result at mild variance with \citet{gill11} 
(hereinafter Paper 1)
and \citet{mcqu12} conclusions.  A primary critique by \citet{basr13}
(hereinafter BWR13)
with undeniable validity, was that CDPP at 6.5 hours of prime relevance
for exoplanet transit detection is not an optimal choice for study of
stellar variability where longer time scales of several days would better
elucidate behavior following from magnetic activity and rotation of 
solar-type stars.

In this paper we revisit the Paper 1 analyses with two 
primary considerations.  First, how do the original conclusions 
regarding noise sources relevant to the detection of exoplanet 
transits change with the consideration of data over 4 years, rather
than the 1.25 years originally used, and with use of data from 
a more mature data processing pipeline providing the time series.
That will be the topic of Section 3.  Second, the topic of Section 4,
will be a consideration of noise for solar-type stars following 
adoption of metrics on a longer time scale of greater relevance to 
primary evidence of magnetic activity induced changes.
Section 5 provides results on simulating the expected
distribution of this longer timescale variability metric.

\section{{\em KEPLER} OBSERVATIONS, DATA RELEASES, AND PRIMARY NOISE METRIC}

Paper 1 provided an extensive discussion of how the 
\eke photometer operated, the selection of targets relevant 
to exoplanet detection, and hence the focus of the noise source study.
Also considered in detail were the primary noise (Poisson from stars,
Poisson from sky background, readout noise, instrument and/or software
imperfections, and intrinsic variability of the stars scaled from 
solar observations) terms expected to be important for CDPP.
Rather than attempting to condense an original three page 
discussion setting the stage for our primary study of noise 
contributions we refer the interested reader to Section 2 of
Paper 1.

The data considered in this paper follow from three epochs:
1) As in Paper 1 the original release of Quarters 2 through 6
in 2009 to 2010.
Quarter 2 was re-released in the middle of this epoch bringing
the treatment of all five quarters to a roughly consistent level.
2) Quarters Q0 -- Q14 as uniformly reprocessed in early 2013.
Quarters 15 - 17 were released at a similar level of software
shortly after this.
3) All quarters as uniformly reprocessed in late 2014.

The early releases of data within three months of 
having been telemetered to the ground used the 
Science Operations Center (SOC) Pipeline 6, with 6.3
being representative of Quarters 2 -- 6 data as
analyzed in Paper 1.  Removal of instrumental systematics,
the key step in producing calibrated \eke time series
was handled via a least squares regression with basis
vectors associated with pointing records, temperature
records, and inferred telescope focus values.
For the early data releases the calibrated data
generated by the Presearch Data Conditioning (PDC) module
for which systematics have been removed was referred
to as {\tt ap\_corr\_flux} in the {\tt fits} files.
Details of processing may be found in the Data Release
Notes applicable to Quarter 5 as a representative 
case \citep{mach10}, the Kepler Data Characteristics
Handbook \citep{chri12b}, and \citet{jenk10}.
While this early software did a good job of removing
instrumental systematics, inspection of light curves
(as discussed in the Data Release Notes) would 
sometimes show clear evidence of spurious signals 
being introduced, as well as frequent removal of likely real
stellar variability.

To address the common suppression of stellar signals a
Baysian approach to PDC was introduced in \eke SOC version 8.0.
This Baysian maximum a posteriori (MAP) approach to 
cotrending \citep{stum12,smit12} more effectively removed
common mode instrumental systematics while preserving
stellar signals.  This earliest version of PDC-MAP was
first applied to Quarter 9 data, as then used in BWR13.

The 2013 data releases were the first time that a uniform reprocessing for the
bulk of \eke mission data was performed.  This used the 
SOC Pipeline 8.3.
The primary change for this data release is that PDC 
uses wavelet decomposition and multiple temporal
scales in performing the MAP processing.  It decomposes
each light curve into three characteristic bands, thus
improving the ability to deal with instrumental systematics,
while still preserving intrinsic stellar signals at 
short to moderate ($\sim$20 days) timescales.  The
longest band ($\geq$21 days) performs a simple robust 
fit to cotrending basis vectors evaluated for this 
temporal band.  Stellar signals at timescales significantly
longer than this may be severely suppressed.
The middle band of 2 hours to 21 days performs a MAP fit.
The shortest band preserves all signals, i.e. no detrending
is performed.  The software evaluates on a star-by-star
basis whether to invoke the multi-scale MAP (msMAP), or
if on the basis of a goodness metric calculated by PDC
regular MAP performs better this is used to provide the
calibrated time series ({\tt PDCSAP\_FLUX}) in the fits file.
About 90\% of the time msMAP is adopted.  Details of 
this processing may be found in \citet{smit12} and
\citet{stum12}, with an update in \citet{stum14}.
Quarters 15 -- 17 were processed by
slightly later versions of the SOC Pipeline, but the changes 
were generally not such as to fundamentally affect
noise characteristics.

The third epoch of data releases in late 2014 considered here has been
the only time that all \eke data were processed
consistently with the same version of the \eke pipeline.
The large change introduced for SOC 8.3 of msMAP was retained.
The primary advance for this newest data release were 
improvements to the lower-level treatment of data at the
pixel level, e.g. a more advanced consideration of overscan
in order to better deal with some of the more serious 
sources of instrumental systematics at a root level.
For details see \citet{thom15}.
This processing used SOC Pipeline 9.2.

\section{STELLAR AND INSTRUMENTAL NOISE DECOMPOSITION}

\subsection{Summary of Original (and Current) Approach}

To facilitate determining the relative importance and quantitative
values of several terms contributing to CDPP we focused on a study 
of a subset of the full \eke sample expected to have comparable 
contributions from the primary terms of simple Poisson fluctuations,
intrinsic stellar variability, and instrument/software imperfections.
By design of the mission \citep{koch10} this led us to focus on
stars of roughly solar-type, and \eke magnitude,
Kp \citep{brow11}, of 12.0 $\pm$ 0.5.
  
We directly modelled contributions of noise from Poisson terms 
on the stellar and sky fluxes, as well as the known CCD readout
noise \citep{chri12b} for each \eke CCD and removed these before attempting to
separate out stellar variability and instrument/software terms.

\eke observations were conducted on the same stellar field, with
primarily the same targets throughout the prime mission.  Four times
during each \eke orbit of the Sun, the spacecraft was reoriented by
90 degrees \citep{vanc09} in order to keep the solar panels illuminated, and
spacecraft radiator in shade.  The progressive reorientation results
in sets of stars cycling through four (of 84 total) CCD channels,
thus providing the primary leverage used to disentangle instrument
and intrinsic stellar contributions.  Considered as an ensemble, if
the noise of one set of stars changes as they cycle through 4 CCD 
channels, then this demonstrates that the electronics associated 
with those channels contribute different levels of noise.
Through adoption of a Singular Value Decomposition (SVD) formalism we
obtained noise terms in time (global value associated with each
quarter as might follow from unique operation of the instrument,
or external factors such as solar particle fluence), space (the 
individual CCD channels), and for the stars.
The SVD formalism follows the discussion in, and uses subroutines
from \citet{pre92} for the solution of a highly over-determined
(more observables than unknowns) set of general linear least-squares
equations with degeneracies present.  A key assumption
was that ensembles of stars nearby on the sky should have the same
intrinsic variability, thus allowing us to put the independently
determined relation of quartets of channels on a common scale.

The original study considered a number of factors such as 
dependence of stellar noise on galactic latitude, crowding of 
sources, and the influence of fainter, superposed background stars.
These proved to be of second order and will not be considered here.
We refer interested readers to Sections 3.1 through
3.8 of Paper 1 for a full discussion of our approach.
In the remainder of this section we focus on results applying 
the SVD formalism as before to updated data products, and the 
use of all 17 quarters of data instead of the original 2 -- 6.

Since four years have passed since the original analysis was 
performed, we started by locating the original codes, recompiling,
and attempting to replicate the sequential analyses of the original
study, using as well the data products used for the 2011 study.
This was successful in that new analyses of the original data
resulted in exactly the results quoted in Tables 1, 2 and 4,
and shown in Figure 8 of Paper 1 giving primary noise separation values.

\subsection{Repeat of Original Updated to New Data Products}

The 2011 study used time series produced within three months 
of the end of each quarter, the last one analyzed (Q6) having
been written in December 2010.  There have since been two 
primary releases in which most, or all of the prime mission data were
reanalyzed with more mature software at the Science Operations 
Center for \ek.  We will provide results separately for the 
processing version 8.3 data released over April through December
2013 for all 17 quarters, and processing version 9.2 released
over November through December 2014 for all the data.

Minor software adjustments needed to be made to accommodate the
newer fits formats of the 8.3 and 9.2 data sets, as well as 
minor modifications in a few cases for date ranges provided in individual
quarters.  With the exception of such details, we have performed
analyses in exactly the same way as in Paper 1.

Adoption of the new data products led to rather dramatic shifts
in the noise levels attributed to individual CCD channels (or 
imperfections in the pipeline software used to analyze them),
as well as dramatic shifts in the noise levels attributed to 
each individual quarter in a global sense.  This was initially a
cause for concern, that perhaps the analyses were either inherently
unstable, or inadequately executed.  The linear correlation of 
variances inferred per channel
between the original study (see Paper 1, Table 2)
and the new one using updated data
products was only $\sim$0.5.  However, examination of the inferred
intrinsic stellar variations between the original data products
for Q2--6, and the newer versions of the same data came in at 
greater than 0.97.  The stars of course had intrinsically the same
behavior independent of how the data were analyzed to remove 
various systematic effects from the time series.  The SVD procedure
successfully returned nearly identical behavior for the stars, 
while showing different and generally smaller noise levels in time
and across the detector channels for the more recently
processed data.

Table 1 shows the assigned quarter-to-quarter excess variance
for the first five full quarters, with the first line being from
Paper 1.  In successive full data releases 8.3 and 9.2
the variance (square of noise) drops dramatically for quarters
2 and 3 which had been most affected by systematics.  This behavior
was expected since most pipeline development after the mission start
was devoted to dealing with and suppressing systematics arising from
imperfection in detector electronics and operational
and environmental variations.

\begin{table}
\begin{center}
\caption{Quarter-to-quarter excess variance.\label{tbl-1}}
\begin{tabular}{cccccc}
\tableline\tableline
Version & Q2 & Q3 & Q4 & Q5 & Q6 \\ 
\tableline
6.3 & 210.46 & 105.82 & 44.52 & 0.00 & 29.89 \\ 
8.3 & 62.35  &   0.00 & 15.02 & 23.15 & 8.64 \\ 
9.2 & 18.23 & 0.00 & 3.01 & 28.46 & 0.04 \\ 
\tableline
\end{tabular}
\tablecomments{Variances in ppm$^2$ over
the five quarters of {\em Kepler} data analyzed.
Variance of quietest quarter is forced to zero.
Version refers to SOC Pipeline version number used.}
\end{center}
\end{table}

Over the three data release versions shown in Table 1 the global 
variance attributed to intrinsic variations of the stars was held
fixed, and as noted above the star-to-star variances were reproduced
at a very high level of fidelity across these.  The mean excess
variance over Quarters 2--6 is 78.1, 21.8 and 9.9 ppm$^2$ over 
data processing releases using SOC 6.3, 8.3 and 9.2 respectively.

We defer showing the individual contributions per channel as in
Table 2, or Figure 12 of the original study until the next section
when data from the full mission are used to set this.  With the 
mean stellar and Poisson contributions held fixed, it is worth noting 
that the mean excess variance from both Table 1, plus the per-channel
excesses drops from 181 ppm$^2$ in the data releases made within 
three months of each quarter end, to 137 ppm$^2$ for release 8.3,
and finally to 98 ppm$^2$ for release 9.2.  Since the sum of stellar
and Poisson terms is 664 ppm$^2$ the component of noise attributable
to imperfections in the detector electronics and the inability of 
detrending software to perfectly compensate has become an increasingly
minor contributor to the overall noise budget, reflecting positive
changes in the pipeline software producing corrected time series.

\subsection{Extension from Quarters 2--6 to full Quarters 1--17}

We have performed full mission analyses for data releases 8.3 and 9.2.
The SVD analysis procedures remain unchanged, but now rather than 
having a nearly minimal solution basis in which each quartet of stars
visited most detector channels only once (with redundancy of quarters
2 and 6), there is now a four-fold redundancy with each set of stars
cycling through the same detector multiple times.  We carry the same
assumption as before, namely that the ensemble properties of the stars
remain fixed in time, and to first order in space as well.  Over a 
four year time span some individual stars are likely to have shown
significant evolution of intrinsic noise within the three-month
quarterly intervals, certainly in going from minimum to maximum 
conditions the Sun shows significant variations.  The SVD solution
relies on having an average of 116 stars per quartet, i.e. the 
individual sets cycling through the detector channels, and it is 
a reasonable assumption that stellar cycle variations are not 
synchronized and the ensemble of 116 stars remains sensibly fixed.
The intrinsic stellar variance star-to-star derived from Quarters 2--6
has a linear correlation of 0.959 with the same as derived from
Quarters 1--17, thus the evolution of intrinsic noise level for 
individual stars is shown to be modest (for the two sets of 9.2
data).

The quarter-to-quarter global excesses are shown in Table 2.
The quietest quarter over 2-16 (the full length quarters) is forced
to zero within the SVD solution, and this happens to be Quarter 9
for both the SOC 8.3 and 9.2 data releases.  The small value shown for 
Quarter 17 is likely an artifact of this being only about one month
long.  The SOC pipeline detrending removes signal on shorter time scales
for this shorter than normal quarter.  The mean of changes over time in the
two independent pipeline processing cases are modest:  50.5 ppm$^2$
on average at 8.3, and 46.6 ppm$^2$ for data release 9.2.  The changes
across time are generally well understood.  High values for quarters
1 and 2 result from a break-in period of less than optimal management
of \eke, e.g. the presence of variable guide stars removed for later
cycles, and multiple safings and repointings in Quarter 2.
Higher values later in the mission, Quarter 12 in particular phase
well with measures of solar activity indicative of increased
particle fluxes encountered by \eke as the Sun transitioned to 
solar maximum activity.
A proxy for what \eke will have experienced is given by the 
Planetary Ap index \citep{sieb71} in Table 2.  This is an average
over measurements of disturbance levels in two horizontal 
field components observed at 13 selected, subauroral stations.
Since \eke was offset by as much as 0.4 AU from the Earth at
the end of mission, an Earth-based metric is only a rough 
indication of the environment at \ek.  These results were 
taken from http://www.solen.info/solar.
Other solar activity indicators such as sunspot number, 10.7 cm
flux, or flare counts also show rising trends with time and 
a good correspondence with the rise in \eke noise in later quarters.

\begin{table}
\begin{center}
\caption{Quarter-to-quarter excess variance.\label{tbl-2}}
\begin{tabular}{rrrr}
\tableline\tableline
Quarter & 8.3 & 9.2 & Ap \\ 
\tableline
1 & 219.06 & 232.03 & 4.53 \\ 
2 & 95.07 & 58.55 & 5.43 \\ 
3 & 3.25 & 9.67 & 2.79 \\ 
4 & 31.65 & 23.85 & 3.48 \\ 
5 & 36.36 & 49.47 & 8.32 \\ 
6 & 20.44 & 16.14 & 7.04 \\ 
7 & 23.84 & 32.46 & 5.00 \\ 
8 & 72.68 & 72.19 & 6.65 \\ 
9 & 0.0 & 0.0 & 8.87 \\ 
10 & 23.56 & 20.41 & 9.68 \\ 
11 & 55.49 & 68.50 & 5.23 \\ 
12 & 125.57 & 127.96 & 11.17 \\ 
13 & 57.53 & 63.78 & 8.90 \\ 
14 & 78.85 & 62.93 & 10.07 \\ 
15 & 78.33 & 44.11 & 6.12 \\ 
16 & 52.92 & 48.44 & 7.53 \\ 
17 & -39.65 & -42.62 & 6.85 \\ 
\tableline
\end{tabular}
\tablecomments{Variances in ppm$^2$
over all 17 quarters of {\em Kepler} data analyzed.  The two columns
are for primary data processing release 8.3 (mid-2013), and 9.2 (late-2014).
Ap is the Planetary A index \citep{sieb71}.}
\end{center}
\end{table}

\begin{figure}
\begin{center}
\includegraphics[width=80mm]{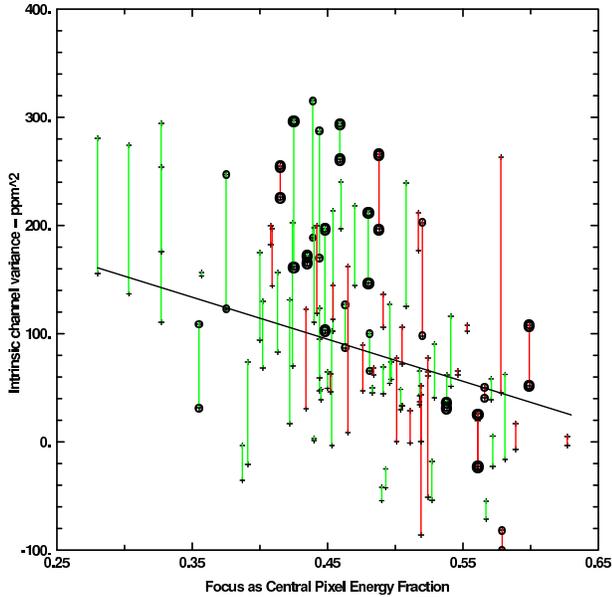}
\end{center}
\caption{By-channel intrinsic variance levels are plotted against
by-channel focus as represented by the fraction of total energy
in the central pixel for a star centered on a pixel.
Channels overplotted with a small circle represent nine cases
independently identified to have moderate Moir\'{e} pattern noise,
and the ten cases with strong Moir\'{e} noise have doubled circles
added.  Values from the original study, and the new 9.2 data release
based on all quarters are both plotted, connected with a green 
line when the new solution has smaller variance, red when larger.}
\label{fig:focnoi}
\end{figure}

The next step in obtaining a separation of error terms between 
the instrument (or residual inability of pipeline software to remove
the results of instrumental imperfections) and stars is to solve
for instrumental terms within each quartet of channels, while at 
the same time solving for the intrinsic variance of each star.
The quartets are then placed on a common scale by requiring that
the ensemble average of stars within each quartet have a common value.
Figure~\ref{fig:focnoi} shows how the new by-channel variances
compare to those found in Paper 1.  For 52 of 84 channels
(62\%) the variance ascribed to the instrument has dropped.
In the original study the channels having poorer focus correlated
strongly with a linear correlation coefficient of -0.63 between 
variance and focus.  In the new set of by-channel variances using
the 9.2 data release and all data as input, this correlation drops
to -0.34.  The correlation of excess noise with poor focus is still
noticeable, however this has been reduced significantly in amplitude.

\subsection{Summary of Changes Using Newer and More Data}

\begin{figure}
\begin{center}
\includegraphics[width=80mm]{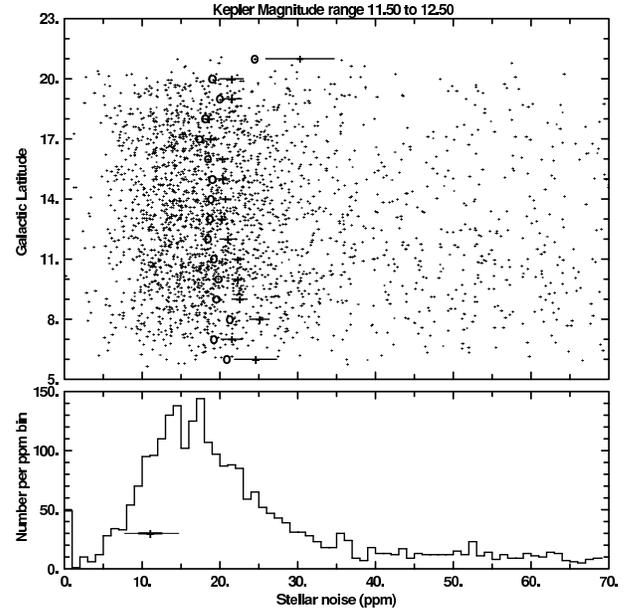}
\end{center}
\caption{Upper panel shows the intrinsic stellar noise in
ppm for the {\em Kp} = 11.5 to 12.5 sample as a function of galactic
latitude.  Medians evaluated up to 100 ppm are shown as `o',
while means from up to 3$\times$ the median at each degree of
galactic latitude are shown as `+' symbols.  Standard errors
for the means are shown.  The lower panel shows a histogram 
of number of stars per ppm bin.  The mean and rms distribution
for solar noise levels over Quarter-long intervals spanning a
solar Cycle are shown by the `+' and heavy horizontal line, 
with the full extent of solar noise per Quarter the thin line.
\label{fig:starint}}
\end{figure}

In repeating the original noise study (Paper 1) using the current
data release (following four years of software development for the
pipeline), and all four years of data we have found generally expected results.
The noise levels attributed to the individual solar-type stars have 
changed very little with adoption of the newer data release; a gratifying
result since pipeline updates cannot have affected the stars.
Figure~\ref{fig:starint} shows the updated version of Figure 8 from Paper 1,
the inferred intrinsic stellar noise, now based on all quarters with 
use of up-to-date pipeline processing inputs.  Only differences of 
minor detail can be noted with respect to the original.
The noise levels inferred for individual channels on the instrument
have dropped with the inclusion of more, and most significantly more
recently processed data.  The software developments within the pipeline
were of course motivated in large part to reduce the excess noise 
attributed to the instrument.  The fraction of variance attributed to 
factors potentially under the control of software development has
dropped from 22\% four years ago, to 13\% now.  This is of course an
over-simplified view.  The importance of changes for various applications
depends not only on a gross measure of noise level, but also on 
detailed characteristics of residual noise.  Similarly the intrinsic 
stellar noise may be amenable to suppression for some applications.
Nonetheless a consistent picture has developed of considerable 
improvement in the pipeline-calibrated data products over time.

\section{USE OF LONGER TIMESCALE NOISE METRIC}

CDPP was designed to capture those components of noise and intrinsic
stellar variability of greatest relevance to the detection of low-amplitude
exoplanet transits having characteristic time scales of 3 to 12 hours.
Such a metric need not be, and indeed is not an optimal one for other studies
such as determining the intrinsic variability of solar-type stars.
The CDPP metric depends on the low frequency tail of variability resulting
from stellar granulation, and only the high frequency tail of variability
resulting from magnetic activity induced variations.  If interested in the
stars it would be better to consider multiple metrics that individually
capture the primary sources of variability. 
The ``flicker", or root mean square variation of stars on
timescales shorter than 8 hours \citep{bast13} has been useful 
for characterizing variability at high frequencies, with resulting
ability to measure stellar gravities, as improved by \citet{kall14}.
At longer timescales the measure of intrinsic stellar behavior is
more difficult given the likelihood of contamination from systematics
in the \eke data.  Once a month \eke suspended science operations
to re-point the fixed high-gain antenna toward the Earth to telemeter
accumulated data to the ground.  This resulted in thermal perturbations
to the telescope and photometer introducing photometric changes large
compared to the stellar variations of quiet solar-type stars.
To deal with these systematics detrending was introduced that very
successfully removed many common mode variations from the instrumental
drifts, but at an additional cost of suppressing true stellar signals
in some regimes and introducing uncertainty in the final product.
Given the roughly month-long rotation period for quiet solar-type 
stars, and the monthly cadence of \eke pointings, recovery
of intrinsic stellar signals on timescales of several days most useful
for characterization of activity variations was thus made challenging.

BWR13 have used two primary metrics to encapsulate stellar
variations on activity timescales.  These are physically well 
motivated, and useful for characterizing stellar activity variations.
Caution, however, is due in application to \eke data where the
pipeline calibration of data may well suppress some variations of relevance
to forming these statistics.  The first diagnostic, $R_{var}$ was 
introduced by \citet{basr11}, this ``range" parameter is found by
sorting all the photometric data points in a given interval (30 days
generally adopted), then taking the difference between the 5\% and 95\%
points (to avoid anomalous excursions) in this distribution.
The resulting statistic will be sensitive to both short timescale
excursions (if lasting more than 5\% of the time during 30 days)
as might follow from star spots, and more generally to longer 
timescale variations approaching the monthly intervals adopted.
Since the \eke systematic noise removal process
(especially the latest msMAP version) strongly suppresses
noise on timescales as short as 30 days, we consider the $R_{var}$
parameter problematic for interpretation of \eke data, especially
when trying to obtain an absolute comparison of placing the Sun
within the distribution of stars assessed with \ek.
The second primary measure of variability introduced in BWR13
is the median differential variability MDV($t_{bin})$.  The MDV measures
the variability by forming bins of length $t_{bin}$, then taking 
the absolute difference between adjacent bins.  The MDV follows as
the median value of the time series of absolute differences.
We have chosen to focus on MDV with $t_{bin}$ = 8 days.  Figure 8 of
BWR13 compares 8-day MDV for the Sun from SOHO data
\citep{froh97}
considering 30 day blocks over a full solar cycle, to an ensemble
of bright \eke stars for one quarter (Q9) of data.  In the comparison
given in their Figure 8 a significant fraction of the solar points 
have MDV values well below the overall minimum value reached
for about 1,000 \eke stars -- an implausible result suggesting
that either something went wrong with evaluating the solar or
\eke values, or that the \eke time series have significant
residuals on time scales longer than 8-days creating this offset.

\begin{figure}
\begin{center}
\includegraphics[width=80mm]{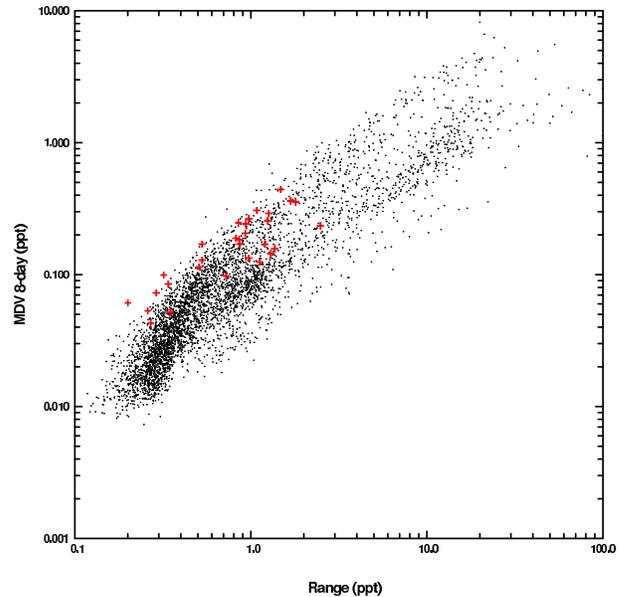}
\end{center}
\caption{Black dots show the median values over all 17 quarters for
the 4,529 \eke solar-type dwarfs for the BWR13 8-day MDV
and Range metrics as computed in this study.  The red crosses show
the same statistics for 30 90-day intervals of solar time series
spanning a full solar cycle.}
\label{fig:mdvsun}
\end{figure}

In order to further pursue a comparison of \eke stars with the Sun
in the 8-day MDV statistic, given the surprising result of many
solar values disjoint from the extrema of 1,000 \eke stars we have
formed our own metrics.  Figure~\ref{fig:mdvsun} shows the 
distribution of solar values for 30 3-month long intervals of SOHO
\citep{froh97}
data, the same set as discussed in Paper 1, and the 4,529
solar-type \eke stars with Kp $<$ 12.5.  Differences with respect
to the BWR13 study include:  (1) We compute 8-day blocks 
over 90 days (or length of \eke quarter of data), rather than within
30 day intervals to form MDV.  (2) We take the median over all 17
quarters of \eke data, rather than adopting Quarter 9.  (3) We use
the latest available (9.2) data release.  None of these differences
are significant.  Our distribution of 8-day MDV values for the Sun
compared to \eke stars is radically different than that in Figure 8
of BWR13.  In particular the distribution of MDV values
for the Sun falls within the extent of stellar values from a large
ensemble of stars.  The radically different distribution follows
primarily from our values for the solar MDV.  Our minimum solar
MDV is about 0.04 ppt, while the BWR13 value is about 
0.0015 ppt.  We differ by over an order of magnitude in scale for
the solar MDV at 8 days.

\begin{figure}
\begin{center}
\includegraphics[width=80mm]{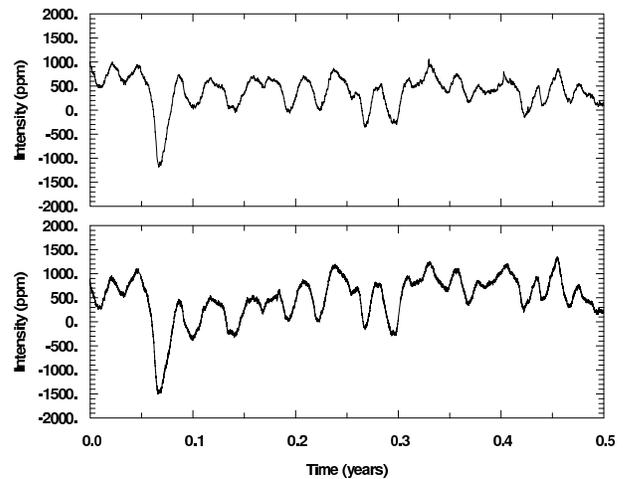}
\end{center}
\caption{Upper panel shows a 0.5 year interval of the solar
activity record adopted by BWR13.  The lower panel
shows the solar record for the same interval as used both in
Paper 1 and this study.}
\label{fig:solboth}
\end{figure}

In order to pursue the latter discrepancy we have compared records
of solar variations used, with the time series used in BWR13
kindly provided by G. Basri.  The latter authors used the mean of
``green" and ``red" VIRGO (SPM) data from SOHO, starting with 
hourly cadence data linearly interpolated to half-hour to roughly
match the \eke cadence.  Paper 1 also used VIRGO/SOHO data,
but started with a compilation at 60 second intervals and binned
this to 29.4 minutes.  We also adopted just the ``green" channel
and scaled this by 0.79 to adjust amplitudes to the longer average
wavelength of \ek.  Figure~\ref{fig:solboth} shows a representative
0.5 year interval between the adopted solar records of the two
studies.  A detailed comparison shows numerous differences, but 
these are at the level of influencing 8-day metrics at the 10\%
level, not the nearly factor of 20 found in our two sets of 8-day
MDV metrics for the Sun.
Our difference from BWR13 for the solar 8-day MDV does not follow
from minor differences in color or sampling for adopted solar records.

Ironically our comparison of solar and \eke 8-day MDV better 
support a primary contention of BWR13 that the 
relative noise levels intrinsic to the stars compared to the
Sun are lower than concluded in earlier studies of Paper 1
and \citet{mcqu12} than does their own result for this metric.
We defer further discussion of typical
activity levels of solar-type \eke stars relative to the Sun
until after discussion of results from our favored long timescale
noise metric.

\subsection{Adoption of a CDPP-style Metric with Longer Timescale}

\begin{figure}
\begin{center}
\includegraphics[width=80mm]{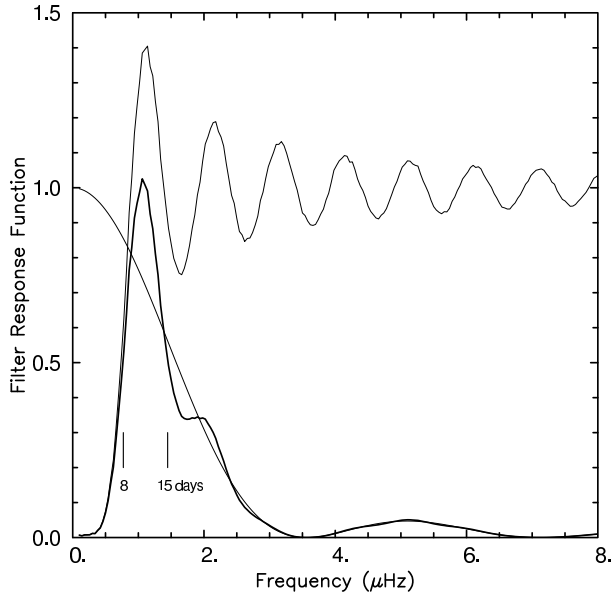}
\end{center}
\caption{Light curves show filter components from the 24 day
quadratic polynomial filtering (zero response at zero frequency),
and sinc function representation of the 3.25 day binning adopted
for the long timescale CDPP.  The bold curve shows the adopted
net response function plotted against frequency with 50\%
transfer periods of about 8 and 15 days flagged.}
\label{fig:filtresp}
\end{figure}

\begin{figure}
\begin{center}
\includegraphics[width=80mm]{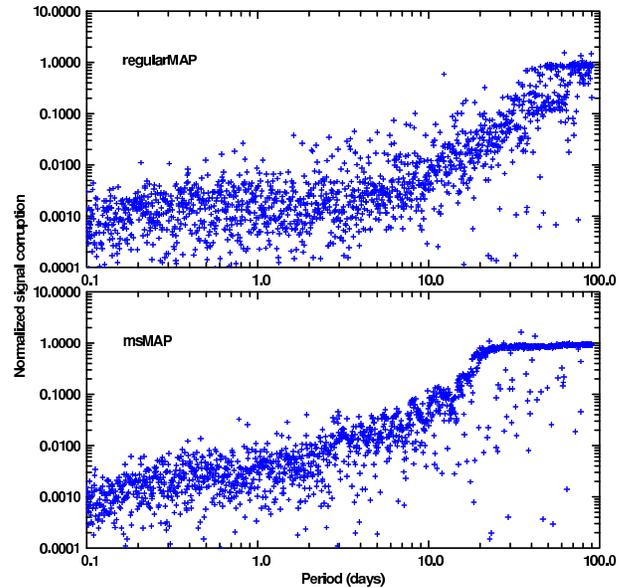}
\end{center}
\caption{This plot shows what fraction of existing signals
in the form of injected test sinusoids at amplitudes corresponding
to one standard deviation of the underlying time series are
preserved by data release 8.0 (`regularMAP' -- upper panel),
and data releases 8.3 and 9.2 (`msMAP' -- lower panel).
Signal corruption of unity corresponds to total loss of signal,
while small values indicate high fidelity retention of input
signals through the systematics removal step.
The corruption metric is qualitatively the same as fractional damping
of the signal amplitude.
Clearly, for the current `msMAP' signals with periods $\geq$20 days 
are severely damped.
Not shown are more subtle, and less well characterized dependencies
on signal amplitude.  Larger input signals show relatively better
preservation at long periods, while smaller amplitudes show more
damping.}
\label{fig:mapdamp}
\end{figure}

The CDPP metric of Paper 1 and in Section 3 above starts
with the calibrated (instrumental signatures removed to the extent
possible) pipeline data, removes a running 2-day quadratic polynomial
fit to the time series, block averages into 6.5 hour intervals, then
evaluates the standard deviation for quarter-long segments.
We choose here to adopt exactly the same procedure, but now use 
timescales longer by $\times$12.  We start with a 24-day quadratic
polynomial fit that will preserve signal at much longer intervals
than the standard 6.5 hour CDPP, then follow this with binning into
3.25 day intervals before evaluating the standard deviation.
Figure~\ref{fig:filtresp} shows the response function of our filtering
and binning operations for this long timescale CDPP. 
The 50\% transfer points are at about 8 and 15 days.
\citet{aigr04} found a timescale of 9.8 days best characterized
solar activity variations, rather than the rotation period
of $\approx$26 days.  Our metric nicely spans the timescale of 9.8 days.
We have also directly verified that the 8 -- 15 day bandpass
represents solar variations at high fidelity by evaluating 
it for 30 ``quarters" of SOHO/Virgo \citep{froh97} data spanning
a full solar cycle.  The 8 -- 15 day bandpass correlates at the 
85\% level with a 8 -- 30 day bandpass measure.  The upper range
of metric being set at 15 days avoids most of the damping inherent
with the calibrated data -- which at 30 days would be largely 
removed, and at 20 days would be uniquely perturbed
star-to-star and quarter-to-quarter.  By design this
timescale was chosen to be as long as possible without the long 
timescale end already having been significantly suppressed with
instrumental systematics removal in the \eke pipeline processing.
Figure~\ref{fig:mapdamp} illustrates the damping introduced
by the pipeline version (``regular MAP", data release 8.0) used
in BWR13, as well as the data products more recently available
(``msMAP", data release 9.2).
This figure supports the selection of an 8 -- 15 day bandpass
filter for our primary long-timescale metric.
This longer timescale CDPP would no longer be relevant to the 
detection of 3 -- 12 hour transits, but is well suited to 
attempting to characterize activity induced variations in a sample
of solar-like stars.

With this much longer timescale metric, and consideration of the 
same stellar sample used in Paper 1 some previously relevant
noise terms are now unimportant.  At 6.5 hours for CDPP the Poisson
noise was roughly comparable to the intrinsic stellar term.
At the $\times${12} longer metric the intrinsic stellar term rises
due to better sampling primary timescales of stellar activity, 
while the Poisson term drops by $\sqrt {12}$.  Factors from readout noise
on the CCDs, Poisson fluctuations on the counts, and sky background
are now unimportant.

\begin{figure}
\begin{center}
\includegraphics[width=80mm]{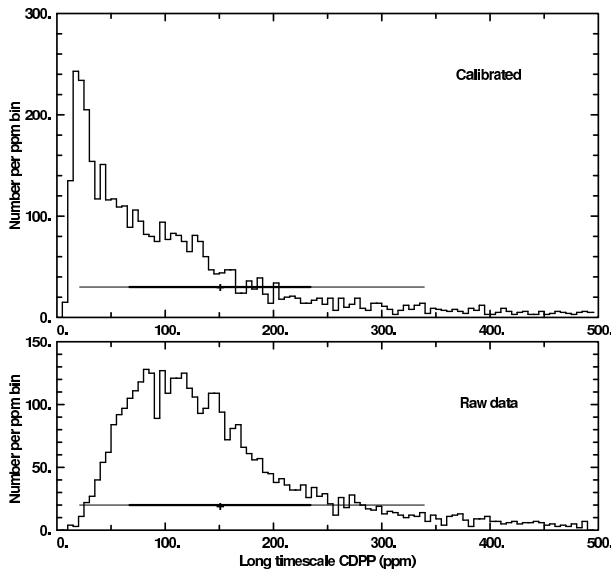}
\end{center}
\caption{The upper panel shows a histogram of number of stars
(of 4,529 total) at different levels of the long timescale CDPP, 
labelled as stellar noise in parts per million based on the
calibrated time series.  The lower panel shows the same based on
use of the direct, or raw data uncorrected for systematics.
The mean and rms distribution for solar noise levels over 
quarter-long intervals spanning a solar Cycle are shown by the
``+" and heavy horizontal line, with the full extent of solar
noise per quarter the thin line.}
\label{fig:cdpplong}
\end{figure}

We have attempted to pursue the same type of Singular Value Decomposition
to isolate noise terms associated with individual quarters, the stars
themselves and contributions from the instrument.  This has been 
relatively unsuccessful.  The original CDPP noise separation leveraged
off isolating nearly comparable terms, and benefited from a relatively
narrow range of intrinsic stellar noise.  The longer timescale CDPP
encounters much more discrepant components in which the instrumental
(or software inadequacy in dealing this these) terms are small compared to 
intrinsic stellar, and more importantly the stars show a broader 
distribution of intrinsic noise.  We therefore concentrate on showing
direct evaluations of the longer timescale CDPP for the \eke stars,
recognizing that if anything these will be over-estimates of the 
intrinsic stellar noise.  We compute the solar metric using the same 
algorithms and codes used for the stars.  The somewhat surprising
results are shown in Figure~\ref{fig:cdpplong}.  Panels are included
for analysis of both the simple aperture photometry (raw), and 
the calibrated data (release 9.2, version for release 8.3 is identical
for all intents) for which the distribution of stellar values (median
for the 17 quarters is adopted for each star) is shown in relation to
statistics on the corresponding solar values.  Note that there is a 
strong cluster in the calibrated data to the low range of solar 
variability.  Even with consideration of the raw-data time series
for which no instrumental systematics have been removed the mode 
for the stars is well below the mean for the Sun.

Great effort has been expended in an attempt to make the primary 
feature (cluster of stellar values to low range of solar) go away.
While one can never be certain of any result, we have been unable
to resolve this finding.  We have verified that our solar record 
in use is reasonable by comparing as in Figure~\ref{fig:solboth}
to an independent compilation.  We have verified that the same 
code is used for the Sun and stars to form the CDPP.
We have verified that the stellar and solar time series are 
normalized in the same way.  Something that could explain the 
upper panel of Figure~\ref{fig:cdpplong} would be significant
suppression of stellar signal within our 8 -- 15 day passband already 
by the \eke pipeline processing.  To pursue this we selected a
subset of stars having CDPP near the mode of 100 ppm in the bottom panel,
that also fell near the much smaller mode near 20 ppm in the upper panel.
We then visually inspected this subset looking for signals
of intermediate frequency (8 -- 15 days) in the raw data that
might have been improperly removed in creating the calibrated data.
While some intermediate frequencies could be seen in the raw
data cases, these invariably seemed to be common mode variations
across the several cases examined, and almost certainly not
inherent stellar signals.  Although expecting that this 
(pipeline suppression of real stellar signals) was the most
logical explanation for the distribution in the calibrated
data of Figure~\ref{fig:cdpplong}, we have been unable 
to find evidence in support of this contention.  Indeed, having
eliminated all potential contenders considered for an explanation we are 
left with accepting the seemingly improbable one that for these
long timescales there is a large subset of the stars having
activity levels near the minimum recently experienced by the Sun.
However, the comparison shown in the upper panel of 
Figure~\ref{fig:cdpplong} is also misleading in over-emphasizing
a quiet distribution for the stars.  To higher CDPP values there
is a very long tail not shown in the figure.  Indeed the number
of stars with CDPP greater than the highest encountered by the
Sun is 802, while the number quieter than the lowest solar value
is only 322.
The mean over all stars is 352 ppm$^2$, while the solar mean is 151 ppm$^2$.
The medians switch to 99 ppm$^2$ for the stars and 163 ppm$^2$ for the Sun.
Recall, though, that we have not made SVD-based
adjustments for other non-stellar contributions to the CDPP.
Although we believe such corrections would be minor at this
long timescale, doing so would not change large values, but 
could shift some of the smaller (at $\lesssim$ 30 ppm)
stellar values to yet lower values.

Table 3 shows the first five lines for the electronically
available table documenting primary results in this paper.
A total of 4529 stars brighter than Kp = 12.5 met the selection
criteria for solar-type dwarfs as detailed in Paper 1.
For each of these stars Table 3 provides the Kp value,
the standard 6.5 hour CDPP analog, and the inferred intrinsic stellar
noise for this based on analysis of all \eke quarters and the 
latest data release as discussed in Section 3.
Also provided are the 3.25 day CDPP raw and calibrated data values
as summarized in Figure~\ref{fig:cdpplong} of this section.

\begin{table*}
\begin{center}
\caption{Standard and long timescale CDPP values.\label{etable}}
\begin{tabular*}{0.6\textwidth}{rrrrrr}
\tableline\tableline
KIC & Kp & CDPP & Stellar Noise & $\times$12 CDPP(raw) & $\times$12 CDPP(cal) \\ 
\tableline
 1025494 & 11.822 &  24.85 &  15.40 &  154.13 &   22.45 \\ 
 1025986 & 10.150 & 119.56 & 118.46 & 5704.62 & 5722.35 \\ 
 1026669 & 12.304 &  28.21 &  17.19 &  244.60 &  105.55 \\ 
 1027030 & 12.344 &  30.40 &  20.27 &  184.49 &  22.97 \\ 
 1162051 & 12.475 &  53.24 &  52.81 &  982.96 &  947.07 \\ 
\tableline
\end{tabular*}
\tablecomments{All noise values are in ppm.  The standard timescale
values for our CDPP analog and inferred intrinsic stellar noise are discussed
in Section 3.  The longer timescale CDPP values corresponding to 
analysis of both raw and calibrated time series are discussed in 
Section 4.  Full version of table is available online.}
\end{center}
\end{table*}

\subsection{Is the Kepler Dwarf Sample More or Less Noisy than the Sun?}

The title of this subsection is a seemingly simple question.
The perhaps best simple answer would be:  It depends.

Previous, careful and reasonable studies that addressed this 
question came up with conflicting answers.  BWR13 and
earlier studies sided with the \eke stars being at least as
quiet as the Sun, while Paper 1 and \citet{mcqu12} 
sided with the \eke stars on average being a bit more active
than solar.

For the long timescale CDPP detailed in this section, one measure
is that more solar-type stars (giants have been excluded) have
variations at a level higher than the most active Sun,
than those having variations
at a level lower than the least active Sun.  However, the mode for
the stellar distribution of activity levels is distinctly toward the
quiet end of the solar range of variability.  This latter feature 
persists were we to adopt our version of the 8-day MDV metric.
This feature would also persist were we to adopt a long timescale
CDPP metric at half the timescale, i.e. with a primary response
function of 4 -- 7.5 days.

Robust removal of instrumental signatures without in some cases
suppressing real stellar signatures is undeniably a difficult problem.
Careful inspection of many raw and calibrated time series suggests
that there is no obvious issue with the pipeline being too 
aggressive and suppressing solar-type star intrinsic variations,
although we cannot fully rule this out as a factor contributing 
to the stellar distribution.

Therefore, perhaps the best answer to settle on is:  We don't know,
it depends.  There probably is not a good, robust answer to the 
simple question posed in this subsection.  What does seem quite 
clear, though, in adopting an answer is that the distribution of
solar variability experienced over a recent solar cycle is well 
within the range that a large number of \eke stars show on average.
The Sun is typical in the \eke distribution, which is quite wide
with a non-simple structure.

\section{SIMULATIONS OF STELLAR NOISE}

In Paper 1 we included extensive discussion of using the
galactic population synthesis package TRILEGAL \citep{gira00}
to provide a simulated set of stars appropriate for the 
\eke field of view.  This was followed by detailed discussion
of granulation and stellar activity contributions, the two of
which were modelled as a function of the TRILEGAL generated
stellar parameters (mass and age).  Normalization was accomplished
for the activity contribution through consideration of both
ground based studies as presented in \citet{radi98}, \citet{lock07},
and \citet{hall09}, as well as reference to solar variations 
as measured by SOHO \citep{froh97}.

With both the simulated stellar parameters and codes available
from the Paper 1 study, we have made only one change:  adoption
of the transfer function shown in Figure~\ref{fig:filtresp} for our $\times$12
longer timescale CDPP metric.  Since for this much longer timescale
metric we expect the stellar contributions at 12th magnitude to 
generally dominate over Poisson, readout noise and instrumental
terms we have provided only the stellar terms from the simulation.

\begin{figure}
\begin{center}
\includegraphics[width=80mm]{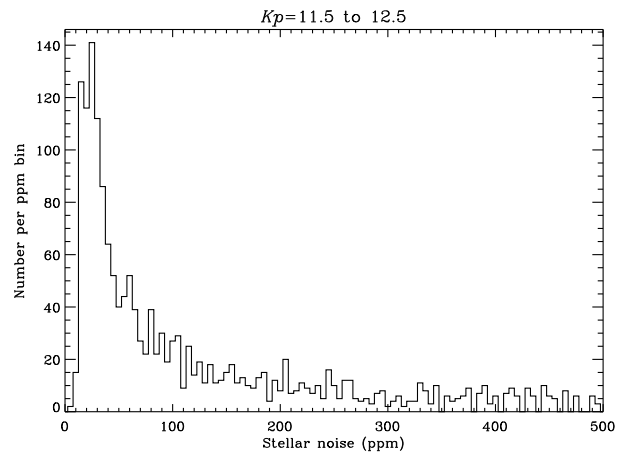}
\end{center}
\caption{Simulated distribution of stellar noise arising from 
activity and granulation for stars with Kp = 11.5 -- 12.5 as 
modelled in Paper 1, with adoption of the transfer function of 
Figure~\ref{fig:filtresp} appropriate to the 3.25 day CDPP metric defined in Section 4.}
\label{fig:simslong}
\end{figure}

Figure~\ref{fig:simslong} shows the resulting distribution of 
simulated stellar noise at the 3.25 day CDPP timescale considered 
in the previous section.  The agreement with observations
as shown in Figure~\ref{fig:cdpplong} is generally quite good.
Stars with parameters close to solar map into mid-range of the 
solar variation as measured directly from the SOHO data.
Most importantly the strong peak at low noise levels -- essentially
a pile-up near the lower range of solar variability levels 
experienced over a solar cycle, is reproduced in the simulations.
Since the simulation codes were not tuned to reproduce the 
distribution seen in the real data of Figure~\ref{fig:cdpplong}, we take the
general agreement as confirmation that the distribution of noise
seen in the real \eke data is a reasonable representation of reality.
The consistency between real and simulated data further
demonstrates that the pipeline is not significantly
suppressing stellar signals in our bandpass.

The population of stars in Figure~\ref{fig:cdpplong} at CDPP
values less than 70 ppm presumably arises from two factors.
The fraction of all stars sampled falling below 70 ppm is 38\%.
The first factor is that $\sim$20\% of the time the Sun 
is this quiet.  The second factor is that 20\% of the stars in
the simulations of Figure~\ref{fig:simslong} have ages greater
than 5 Gyr.  Thus the very quiet stars sampled by \eke may
arise equally from stars similar to the Sun, and in quiet 
phases of activity cycles, and from stars inherently older
than solar.

\section{Summary}

We have repeated an earlier analysis studying noise in \eke data
at timescales relevant to the detection of exoplanet transits
using much longer time intervals, and making use of more 
recent data products.  The inferred intrinsic stellar noise stayed
fixed with adoption of more, and newer data, thus providing 
confidence in the analyses.  The inferred residual noise arising
from the instrument dropped with the consideration of newer data
products, this of course would be expected since the software 
updates had been intended to do this.  Residual noise as a function
of time during the \eke mission correlates well with solar activity.
The earlier study by us (Paper 1) had shown a strong correlation
between excess noise by-channel with the mean focus offset of the 
channels (in the sense that fuzzier images had poorer photometry).
That correlation is still present considering all of the data,
and the most recent data release, but is now relatively weak, consistent
with most possible gains in suppressing instrumental noise now being
in hand.

We have explored a longer timescale metric better suited to elucidating
levels of stellar magnetic activity induced variations.
This has shown mixed results.  We find that the spread of solar
variations over a recent cycle are well within the spread of mean
noise levels for a large sample of solar-type stars.  We also 
find that there is a strong concentration of \eke noise levels
near the minimum values reached by the Sun.  We have not been 
able to find evidence in support of any conclusion for this, 
except the one directly presented:  there seem to be many \eke
solar-type stars that are as quiet as the quiet Sun, more than 
we expected based on either the earlier (Paper 1) study
using a metric less well suited to characterizing stars, or to
modelling of expected noise levels using galactic population 
synthesis models \citep[][TRILEGAL]{robi03} suggesting an age distribution averaging younger
than the Sun.  A direct simulation for the long timescale does,
however, show results consistent with the observations.
A significant fraction of stars are older, and hence quieter
than the Sun even though as argued in Paper 1 the overall age
distribution is younger than solar.
As such this study shows that the Sun may be
considered typical of the \eke distribution of solar-type star
activity levels.
The significant fraction of stars with activity levels at
or below the quiet Sun is a generally positive result for
habitability \citep{see14} of potential Earth-analogs in 
the \eke field.
A simple answer to the question of whether the
Sun is quieter or noisier than the \eke sample has not been reached.

\acknowledgements 

Funding for \ek, the tenth Discovery mission, was provided by NASA's Science Mission Directorate.
The many people contributing to the development of the {\em Kepler Mission}
are gratefully acknowledged.
We thank Joseph Twicken for shepherding evolution of the 
\eke data processing pipeline.
R.L.G. and L.W.R. have been partially supported through grant
NNX14AK65G-S01 of the NASA ADAP program.
The Center for Exoplanets and Habitable Worlds is supported by the 
Pennsylvania State University, the Eberly College of Science, and 
the Pennsylvania Space Grant Consortium.
Data presented in this paper were obtained from the 
Mikulski Archive for Space Telescopes.
Support to MAST for non-{\em HST} data is provided by the NASA
Office of Space Science via grant NNX13AC07G and by other
grants and contracts.

{\it Facilities:} \facility{{\em Kepler}}.

\end{document}